# Unidirectional light transmission by two-layer nanostructures interacting via optical near-fields

MAKOTO NARUSE[1*], SATOSHI ISHII[2], JEAN-FRANCOIS MOTTE[3], AURÉLIEN DREZET[3], SERGE HUANT[3], AND HIROKAZU HORI[4]

[1] National Institute of Information and Communications Technology, Tokyo, Japan

[2] National Institute for Materials Science, Tsukuba, Japan

[3] Univ. Grenoble Alpes, CNRS, Institut Néel, 38000 Grenoble, France

[4] University of Yamanashi, Kofu, Japan

*Corresponding author: naruse@nict.go.jp

**ABSTRACT**

**We experimentally demonstrate unidirectional light transmission through two-layer nanostructured materials considering that the horizontal-to-vertical-polarization conversion efficiency in the forward direction and the vertical-to-horizontal efficiency in the backward direction, which are usually equivalent due to optical reciprocity, are different. The different ways of transferring light momentum in the forward and backward directions via optical near-fields between the layers are responsible for the unidirectionality of light, which was theoretically investigated in our recent work [J. Opt. Soc. Am. B 31, 2404-2413]. With two-layer metal nanostructures experimentally fabricated via a repeated lift-off technique,**

**consistent optical characteristics are observed, verifying the utilization of the large momentum of optical near-fields.**

Unidirectional light transmission is important for various applications such as optical isolators and one-way mirrors. Asymmetric transmission of linearly or circularly polarized light by optical near-field processes on the nanometer scale through planar nanostructures has been intensively studied[1-10] as well as nanostructured waveguide-based architectures.[11-13] In Ref. 1, we theoretically demonstrated that optical near-fields between two-layer planar nanostructures transfer light momentum in a different manner between forward and backward directions based on the angular spectrum representation of optical near-fields (Fig. 1(a)).[1,14] Utilization of large momenta available in the optical near-field is one of the most important and unique aspects of nano-optics.[15] The polarization conversion efficiency from $x(y)$-polarized input light to $y(x)$-polarized output light in the forward direction (denoted as $T^{(F)}_{X \to Y}$ and $T^{(F)}_{Y \to X}$) differs from the polarization conversion efficiency from $y(x)$-polarization input to $x(y)$-polarization output in the backward direction ($T^{(B)}_{Y \to X}$ and $T^{(B)}_{X \to Y}$), meaning that the optical reciprocity does not hold for the particular polarizations in either horizontal or vertical directions. Such a property is called unidirectional light transmission.[1]

In this study, we provide a proof-of-principle, experimental demonstration of unidirectional light transmission by fabricating gold (Au)-based two-layer nanostructures embedded in $SiO_2$ substrates using electron-beam (e-beam) lithography and repeated lift-off processes. Expected unidirectional optical transmission characteristics are observed. Here we describe the fabrication of the two-layer nanostructure, followed by its optical characterization.

Based on the previous work[1], unidirectionality is expected in the near-infrared regime when the first-layer nanostructure comprises 80-nm-wide gratings and the second layer consists of pairs of L-shaped structures with a side length of 200 nm, as schematically shown in Fig. 1(b). The *x*- and *y*-axis indicated therein correspond to the horizontal and vertical directions of polarizations, respectively, when they are subjected to input light.

The device was fabricated by applying lift-off process twice. The steps involved were as follows:

[STEP 0] We first fabricated a centimeter-to-micrometer-scale, large-sized Au pattern on $SiO_2$ substrate by laser lithography in order to clearly indicate the zone of interest (ZOI) for the subsequent processes.

[STEP 1] By utilizing the ZOI alignment marks made in STEP 0, the first e-beam lithography was conducted to fabricate the L-shaped structure. A 100-nm-thick Au layer was deposited on top of a 4-nm-thick titanium (Ti) layer. Fig. 2(a) shows a scanning electron microscope (SEM) image of the fabricated structures after the first lift-off where clear L-shapes are observed.

[STEP 2] After depositing $SiO_2$ to obtain a 200–300 nm gap as the interlayer spacer material between the first and second layers, 4-nm-thick Ti and 100-nm-thick Au layers were formed on top of the device after STEP 1. Then, the second e-beam lithography was performed to fabricate 80-nm-wide gratings with the help of the ZOI markers made in STEP 0.

[STEP 3] An additional $SiO_2$ layer was deposited to cover the Au structures of the second layer. A cross-sectional viewgraph of the intended design is schematically shown in Fig. 2(b).

After the omission of the deposition of the interlayer $SiO_2$ in STEP 2, the SEM images of the resultant structure when the grating structures are additionally formed after fabricating the L-shapes in STEP 1, are shown in Figs. 2(c) and (d). Clear gratings and L-shapes are observed in Fig. 2(c) while the magnified image in Fig. 2(d) manifests the degradation of the design; for example, the linewidths range from 86 to 97 nm (as opposed to 80 nm in the design). However, it should be emphasized that the alignment of the grating structure (by the second e-beam lithography) to the L-shaped structure (by the first e-beam lithography) was performed with nanometer-level accuracy. Meanwhile, a thin Ti layer was added on top of the structures shown in Figs. 2(c) and (d); this was done for clearly evaluating the on-surface loads of the sample by the incidental beam of the SEM, whereas such a Ti layer was not prepared on the structures shown in Figs. 2(a) and (e). We did not use an atomic layer deposition (ALD) technique for fabricating the $SiO_2$ interlayer between the two Au layers; this is because our ALD machine is not equipped with a $SiO_2$ precursor, as well as due to technical difficulty, including the long duration required for the formation of 200–300-nm-thick $SiO_2$ layer by ALD.

Fig. 2(e) shows an SEM image of the resultant two-layer device including the deposition of $SiO_2$ in STEP 2 and 3. Although the presence of $SiO_2$ layers inhibits clear imaging of the fabricated Au structures, the shapes of the first layer structures (L-shapes) were deformed through the process in STEP 2. Nevertheless, in this first proof-of-principle experiments, we proceeded with optical characterization while recognizing the fabrication difficulties, as discussed at the end of the paper.

In the optical characterization, the device was irradiated with white light (halogen lamp) via an objective lens (50×, NA0.8), and the transmitted light was collected by another objective lens (60×). Two polarizers (extinction ratio: $10^3$) were arranged in a crossed-Nicols manner. We defined the *forward* light propagation by setting the nanostructured patterns to face the white light, whereas the *backward* light propagation was defined when the $SiO_2$ substrate-side of the device faced the white light. The transmitted output light intensity was measured by a spectrometer (ACTON SP2300, Princeton Instruments) equipped with CCD (DU401A, Andor).

As described in the beginning, unidirectionality is evaluated as the difference between the polarization conversion efficiencies in the forward and backward directions. The red curve in Fig. 3(a) indicates the polarization conversion efficiency from *x*-polarized input light (abbreviated as *x*-IN hereafter) to *y*-polarized output light (*y*-OUT) in the *forward* direction, denoted as $T_{X \to Y}^{(F)}$. Here, the moving-averaged values spanning 10 data points in the spectra, corresponding to a wavelength of approximately 5 nm, are displayed. Likewise, $T_{Y \to X}^{(F)}$ depicts the conversion efficiency from *y*-IN to *x*-OUT in the forward direction, as denoted by the green curve. Similarly, $T_{X \to Y}^{(B)}$ and $T_{Y \to X}^{(B)}$ show *x*-IN to *y*-OUT and *y*-IN to *x*-OUT polarization conversion efficiencies in the *backward* direction marked by the magenta and blue curves, respectively.

In the optical experiments, we needed to flip the device manually in order to examine the transmissions in the forward and backward directions, which causes unavoidable imperfection resulting from optical misalignment. To compensate for such effects, we conducted the following normalization. Usually, $T_{X \to Y}^{(F)}$ ($T_{Y \to X}^{(F)}$) is equal to $T_{Y \to X}^{(B)}$ ($T_{X \to Y}^{(B)}$) based on optical reciprocity. However, the two-layer nanostructures of our design generate non-equality between these figures due to the different way of momentum transfer via the optical near-field.[1] We assume that the equality holds at the shortest wavelength of our examination, which is approximately 830 nm.

The unidirectionality is examined by the non-equalities of these polarization conversion efficiencies; $|T_{X \to Y}^{(F)} - T_{Y \to X}^{(B)}|$ and $|T_{Y \to X}^{(F)} - T_{X \to Y}^{(B)}|$ are shown by the red and blue curves in Fig. 3(b), respectively. Further, Figs. 3(c) and (d) summarize the numerically calculated polarization conversion efficiencies and the unidirectionality investigated in Ref. 1, respectively.

The two curves in Fig. 3(b) exhibit large values in the wavelength regime greater than 1000 nm, which is consistent with the numerical considerations (Fig. 3(d)). However, the detailed characteristics exhibit differences between the experiments and simulations; for example, $|T_{Y \to X}^{(F)} - T_{X \to Y}^{(B)}|$ (blue curve in Fig. 3(b)) exhibit greater values in the long-wavelength regime around 1300 nm, whereas simulations indicate smaller values throughout the spectrum (blue curve in Fig. 3(d)). This can be attributed to the differences between simulation and experiment in terms of both ~~exact~~ device structure and optical setup.

Nevertheless, even though a quantitative agreement between experiment and simulation is presently out of reach, the results indicate a consistent tendency between the experiment and numerical design studies, which verifies the concepts and principles of unidirectionality originating from near-field interactions between two layers, reported in Ref. 1.

Finally, we note a few remarks for future studies. As indicated by the SEM image in Fig. 2(e), structural degradations are observed in the fabricated two-layer devices, whereas fine nanostructures are fabricated for the one-layer structure, with the alignment having nanometer-scale precision in spite of repeated e-beam and lift-off processes. This indicates that a technological improvement is necessary for multilayer nanostructured materials. Indeed, for example, Yao *et al.* utilized nanoimprinting technologies for layered grating structures.[6] Self-organized fabrication approaches, such as the one based on DNA-origami, are of interest, as demonstrated by Wang *et al.* for a honeycomb structure.[16] Meanwhile, realizing further functionality from the material aspect is interesting. For example, Nakagomi *et al.* demonstrated nanometer-scale near-field optical control of photochromic materials[17]; shape-engineered photochromic materials will provide additional intrinsic attributes. Moreover, the grating structure on the second layer can provide the ability for electrical modulation via an externally connected device if the material is conductive, leading to the reconfigurability of its optical properties. Oxide materials, such as $VO_2$ (recently used in metasurface studies[18]), are potential candidates for future developments.

In summary, with experimentally fabricated two-layered metal nanostructures, we examine unidirectional light transmission in the sense that polarization conversion efficiencies in the forward and backward directions, which are normally the same because of optical reciprocity, are different. The experimentally observed optical transmissions are consistent with the theoretically and numerically predicted ones in Ref. 1. The results verify the successful utilization of the difference in the momentum transfer carried by optical near-fields between forward and backward directions.

**Acknowledgment**

The study is supported in part by Core-to-Core Program A. Advanced Research Networks, Grants-in-Aid for Scientific Research (A) (JP17H01277) from Japan Society for the Promotion of Science; CREST program (JPMJCR17N2) from Japan Science and Technology Agency.

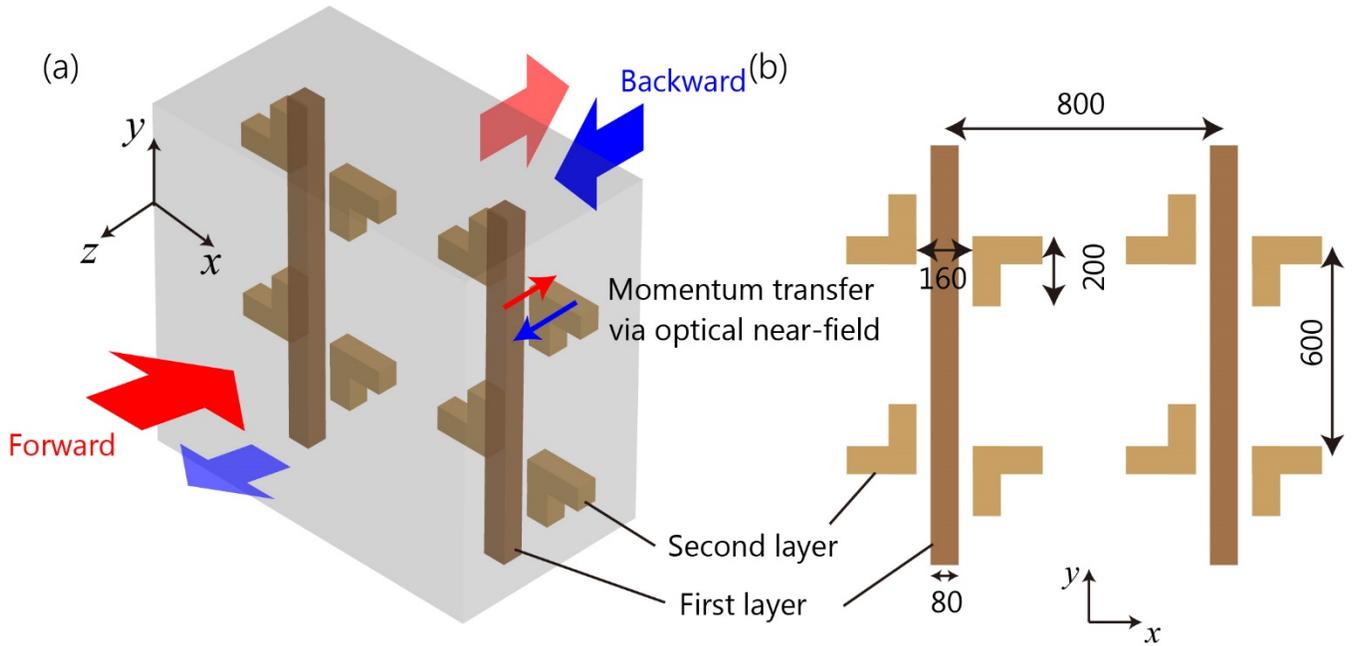

**Fig. 1.** (a) Optical near-fields between two-layer nanostructures transfer light momentum in a different manner between forward and backward directions, leading to unidirectional light propagation.[1] (b) Design of the two-layer nanostructure.

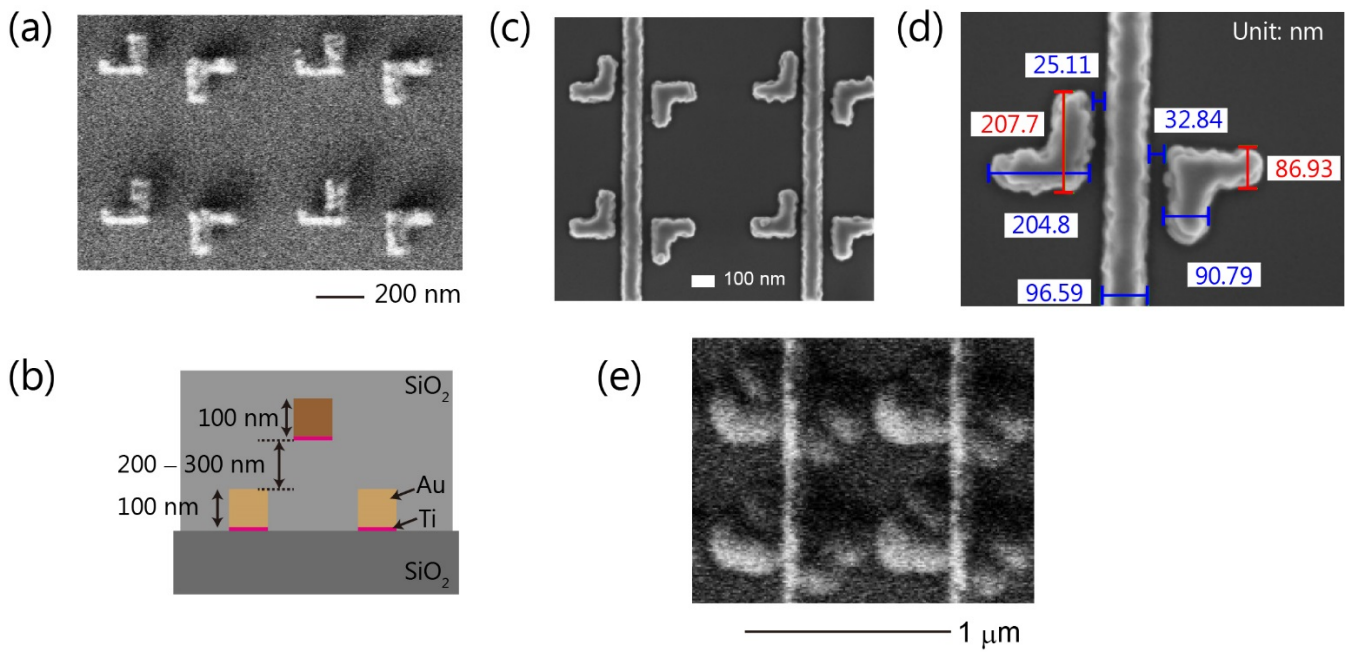

**Fig. 2.** (a) L-shaped structures fabricated via first e-beam lithography and lift-off. (b) A schematic diagram of the horizontal cross-sectional profile of the proposed device. (c,d) After (a), the grating structures were fabricated via second e-beam lithography. The alignment between the first and second lithography/lift-off was achieved with nanometer-scale precision. (e) Fabricated two-layer structures containing $SiO_2$ in the interlayer and cover layer.

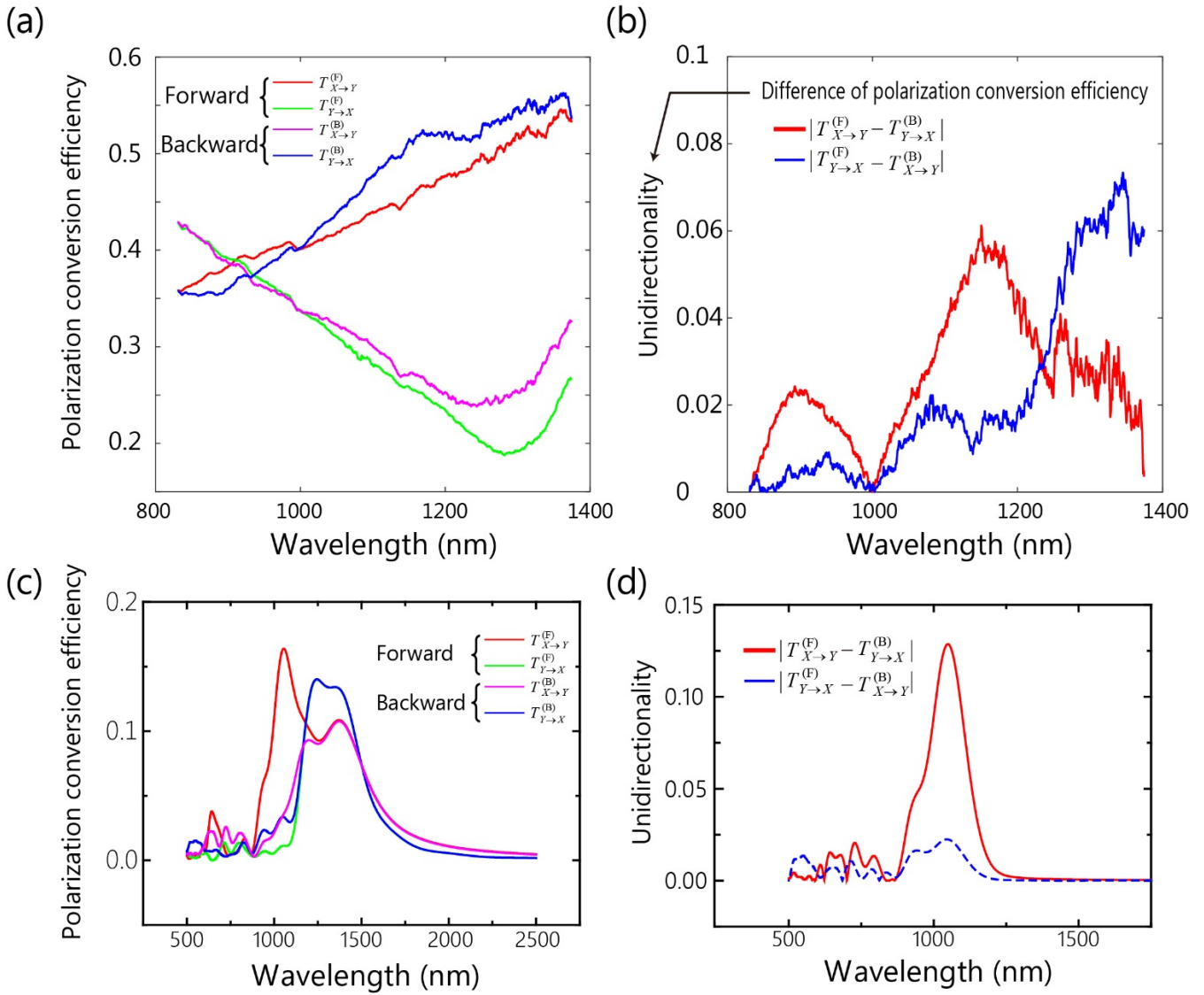

**Fig. 3.** (a) Experimental results of polarization conversion efficiency, and (b) graph of unidirectionality given by the absolute difference between the polarization conversion efficiencies in forward and backward directions. (c,d) Numerically computed polarization conversion efficiencies and unidirectionality adapted from Ref. 1.